\documentclass[prd,nofootinbib,floatfix,superscriptaddress,showpacs,twocolumn,amssymb,amsmath,amsfonts,aps]{revtex4}

\usepackage{graphicx} 
\usepackage{dcolumn}  
\usepackage{longtable}
\usepackage{bm}       
\usepackage{subfigure}
\usepackage{multirow} %
\usepackage{sidecap} 
\usepackage{rotating}
\usepackage{longtable}
\usepackage{bbold}
\usepackage{color}
\input{babarsym}
\usepackage{hyperref}    
\usepackage[all]{hypcap} 
\begin{document}
\author{Biplab Dey}
\affiliation{Sezione INFN di Milano, Milano, Italy}
\date{\today}

\title{Angular analyses of exclusive $\Bbar \to X_J \ell_1 \ell_2$ decays for spin $J \leq 4$}

%
%
\begin{abstract}

As an update to our previous calculation for spin $J\leq2$, we present the angular moments for exclusive $\Bbar \to X_J \ell_1 \ell_2$, where $\ell_1$ is a charged massless lepton and $\ell_2$ is a charged or neutral massless lepton, and $X_J$ is a mesonic system with spin $J\leq4$. The expected applications include higher resonances in the $[K\pi]$ system in $\Bzb \to \Km \pip \mun \mup$ at LHCb in Run~II, and in the $[\pi \pi]$ system for $\Bbar \to \pi \pi \ellm \barnuell$ at Belle~II. For the $J\leq2$ case, we also provide a set of consistency relations among the measured moments observables and validate them against the latest measurements from LHCb.

\end{abstract}

\maketitle

\section{Introduction}
\label{sec:intro}

With the advent of the next generation of $B$ factories at LHCb and Belle~II, hitherto rare decays are quickly shedding the ``rare'' tag. A good example is the flavor changing neutral current decay $\Bzb \to \overline{K}^{\ast 0} \mun \mup$~\cite{LHCb-PAPER-2015-051}, where the signal yield is expected to reach $\mathcal{O}(10^4)$ after Run~II at LHCb. It is therefore pertinent to look beyond the dominant hadronic resonant structures contributing to rare or Cabibbo-suppressed $B$ decays in the electroweak penguin and semileptonic sectors, which, together we classify as $\Bb \to X_J \ell_1 \ell_2$. The hadronic $X_J$ system here is typically $[\pi \pi]$ or $[K\pi]$ and the leptons $\ell_{1,2}$ are assumed massless. The goal of this paper is lay out the formalism for performing angular analyses incorporating these higher spin states, from an experimentalist's perspective. The spin $J \leq 2$ case was addressed in a previous work~\cite{Dey:2015rqa} and subsequently employed to to analyze LHCb Run~I~\cite{lhcb_spd} data in the $\Bzb \to \Km \pip \mun \mup$ mode, with $\qsq\in[1.1,6.0]$~GeV$^2$ and $\mkpi \in [1330,1530]$~MeV, where $S$-, $P$- and $D$-waves occur in the $[K\pi]$ system. Here $\sqrt{\qsq}$ is the invariant di-lepton mass. The interesting feature of the new LHCb data~\cite{lhcb_spd} is that it indicates a non-dominant $D$-wave contribution, contary to expectations from both existing $\Bzb \to \jpsi \Km\pip$~\cite{Chilikin:2014bkk} and $\Bzb \to \Km \pip \gamma$~\cite{Nishida:2002me,Aubert:2003zs} data and previous theory work~\cite{Lu:2011jm}, where the $K^\ast_2(1430)$ state plays a dominant role in the $\mkpi \sim 1430$~MeV region. The new results could point toward a revised understanding of the underlying form-factors.

The present paper extends the relevant observables to spin $J \leq 4$, exhausting to a high degree, the known resonant structures listed in the PDG~\cite{Agashe:2014kda}. The motivation to look at these higher spin states is the richer spectrum of angular observables they offer. Belle has already observed the decay $B^- \to f_2(1270) \ellm \barnuell$~\cite{Sibidanov:2013rkk} while \babar\; has probed semileptonic $B$ decays to excited $D^{\ast \ast}$ states~\cite{Lees:2015eya}. Even higher spin structures such as $K^\ast_4(2045)$~\cite{Lu:2011jm} are expected to be accessible during the ongoing Run~II data-taking period at LHCb. 

In addition, we also investigate the issue of the measured angular moments observables not being independent variables. This results in a set of consistency relations among them. For the $J\leq2$ case, we provide eight of these relations and validate them against the latest measurements from LHCb~\cite{lhcb_spd}.

\section{The angle conventions}


Figure~\ref{fig:angle_conventions}a shows the three concerned angles for the prototypical $\Bb \to X \ell_1 \ell_2$ decay, using the $\Bzb \to \Km \pip \ellp \ellm$ mode. The hadronic- and leptonic-side helicity angles are $\thetav$ and $\thetal$, respectively, while $\chi$ is the dihedral angle between the hadronic and leptonic decay planes. 
For three-body decays on the hadronic side, such as $\omega\to 3\pi$, the normal to the decay plane defines the analyzing direction. Following Ref.~\cite{Gilman:1989uy}, in the mother $\Bb$ rest frame, the back-to-back leptonic and hadronic systems share a common $\hat{y}$ axis, with opposite $\hat{x}$ and $\hat{z}$. For the $\Bb$ (containing a $b$-quark) in Fig.~\ref{fig:angle_conventions}a, we follow the negatively charged lepton to define the leptonic helicity angle $\thetal$. The hadronic helicity angle, $\thetav$ is defined similarly. The quadrant of the dihedral angle $\chi$ is fixed, by fixing the azimuthal angle of the $\ellm$ to be zero in the leptonic helicity frame. The azimuthal angle of the hadronic side analyzer in the hadronic helicity frame then defines $\chi$. 

\subsection{CP conjugation}

\begin{figure}
\centering
\subfigure[]{
\centering
\includegraphics[width=0.45\textwidth]{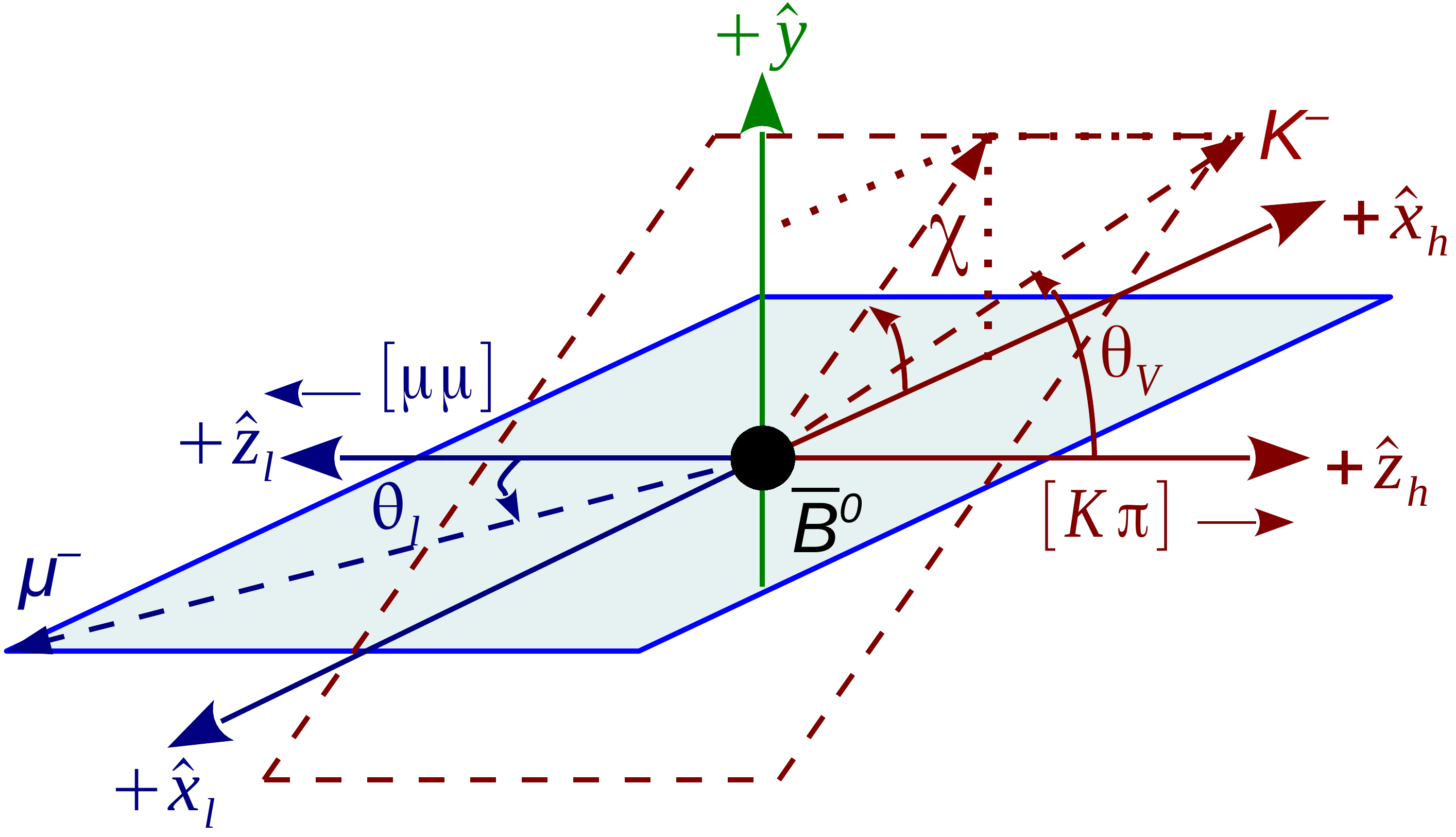}
}
\subfigure[]{
\centering
\includegraphics[width=0.45\textwidth]{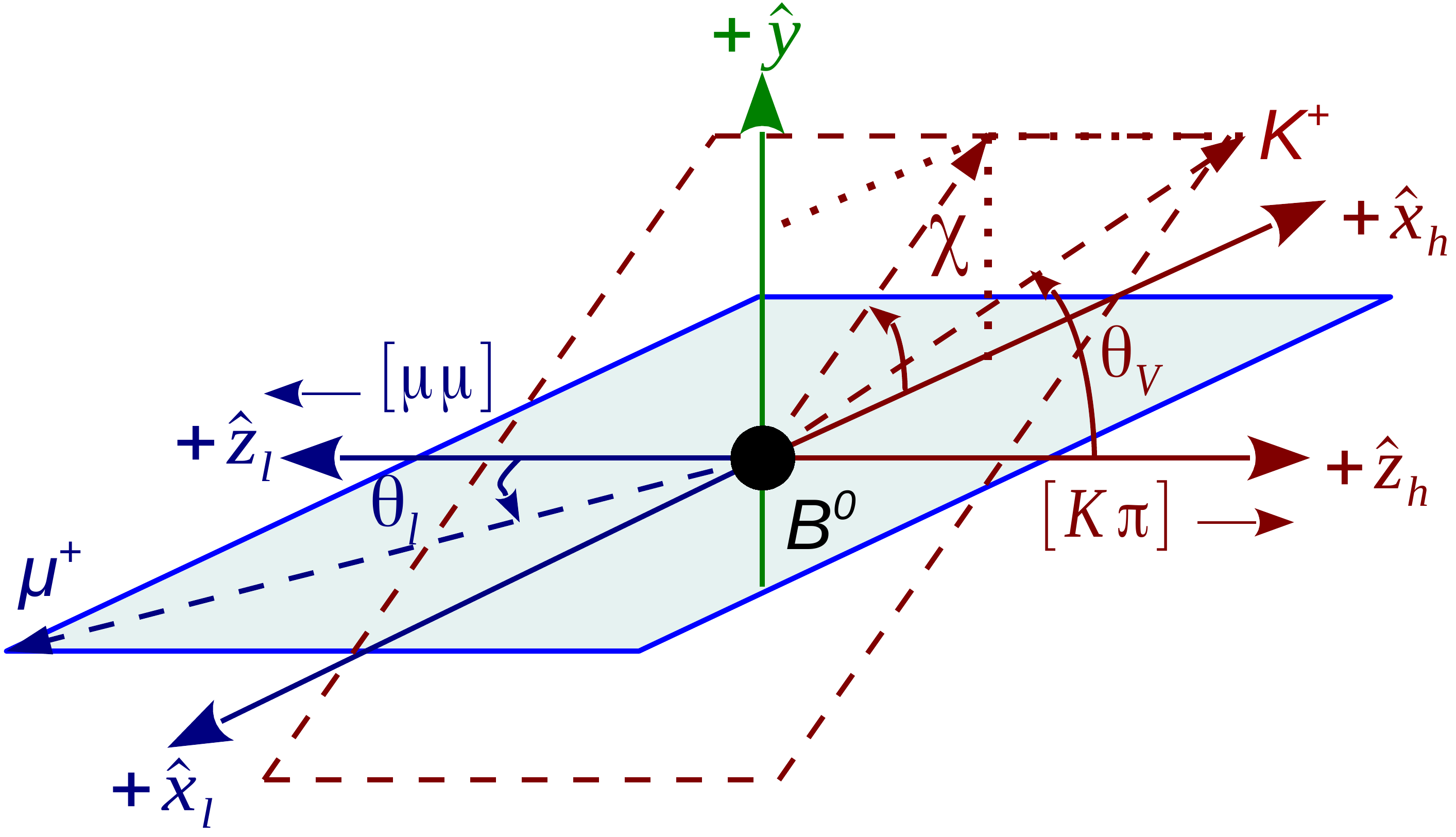}
}
\caption{Angle conventions for the (a) $\Bzb \to \Km \pip \mun \mup$ (b)  $\Bz \to \Kp \pim \mup \mun$. This serves the prototypical  $\Bb \to X \ell_1 \ell_2$ topology in both semileptonic and electroweak penguin analyses. The leptonic and hadronic frames are back-to-back with a common $\hat{y}$ axis.}
\label{fig:angle_conventions}
\end{figure}

For the CP conjugated decay $B \to \overline{X} \overline{\ell}_1 \overline{\ell}_2$ in Fig.~\ref{fig:angle_conventions}b, we follow the charge-conjugated particles. That is, if we followed the $\mun$/$\Km$ for the $\Bb$, we follow the $\mup$/$\Kp$ for the $B$. This is shown in Fig.~\ref{fig:angle_conventions}b. The expressions of angular observables in terms of its corresponding amplitudes remain the same during this procedure. However, the amplitudes themselves are related by~\footnote{The conjugation relations for CP eigenstates $B^0_{(s)} \to h^+ h^- \ellp \ellm$ are different and include additional mixing terms.}
\begin{align}
\mathcal{H}^\eta_\lambda(\delta_W,\delta_S) = \overline{\mathcal{H}}^{-\eta}_{-\lambda}(-\delta_W,\delta_S),
\end{align}
where $\eta = -1 (+1)$ denotes the leptonic-side left(right)-handed amplitudes and $\delta_W$($\delta_S$) are the weak(strong) phases. In the absence of direct CP violation, it can be checked that the measured observables get a sign flip for the terms odd in $\chi$. For convenience, experimentalists flip the sign of $\chi$ during the CP conjugation. This way, in the absense of CP violation, the measured observables are the same between the $\Bb$ and $B$ decays and one can conveniently merge the two datasets for the CP-averaged measurements.

It is important to note that while reporting the helicity amplitudes, it is pertinent to explicitly mention whether the amplitudes correspond to the $\Bb$ or the $B$. The helicity tags in the amplitudes are dictated by the underlying couplings, and are not convention-dependent. For example, in semileptonic decays the $(V-A)$ structure ensures that $|H_-| > |H_+|$ for the $\Bb$, involving an underlying $b$-quark transition.

\subsection{Conversion dictionary for semileptonic decays}

The three commonly followed theory frameworks for semi-leptonic decays are Gilman-Singleton (GS)~\cite{Gilman:1989uy}, Korner-Schuler (KS)~\cite{Korner:1989qb} and Richman-Burchat (RB)~\cite{Richman:1995wm}. The charged lepton is always used to define $\thetal$. We follow the GS convention in this paper, while the KS convention is mostly followed in semi-taonic analyses. The translation between the three angles for the $\Bb$ decays is given in Table~\ref{tab:sl_dict}.

\begin{table}
\begin{tabular}{c|c|c}
Angle & KS & RB \\ \hline 
$\thetal$ & $\pi - \theta^{\rm GS}_\ell$ &  $ \theta^{\rm GS}_\ell$ \\ 
$\thetav$ & $\theta^{\rm GS}_V$ &  $ \theta^{\rm GS}_V$ \\ 
$\chi$ & $\pi+\chi^{\rm GS}$ &  $ \pi + \chi^{\rm GS}$ 
\end{tabular}
\caption{Translation dictionary between GS~\cite{Gilman:1989uy}, KS~\cite{Korner:1989qb} and RB~\cite{Richman:1995wm} conventions for the semileptonic $\Bb$ decay.}
\label{tab:sl_dict}
\end{table}

\subsection{Conversion dictionary for electroweak penguin decays}

For electroweak penguins, there are existing LHCb $\Bzb\to \overline{K}^\ast\mun\mup$ data~\cite{Aaij:2013iag,Aaij:2015oid} with one particular convention. The erstwhile theory community~\cite{Egede:2010zc} used a different convention. The situation is further complicated because, recently, theory papers have appeared~\cite{Becirevic:2016zri, Gratrex:2015hna} that use the LHCb $\Bzb \to \overline{K}^\ast\mun\mup$ conventions. With reference to Fig.~\ref{fig:angle_conventions}, the angle conventions for electroweak penguins are listed in Table~\ref{tab:ewp_dict}. The negatively charged lepton is used to define $\thetal$ for the $\Bb$ decay. We also note that the latest LHCb $\Bzb \to \Km \pip \mun\mup$~\cite{lhcb_spd} paper follows the GS conventions.

\begin{table}
\begin{tabular}{c|c|c}
Angle & LHCb $\Bzb\to \overline{K}^\ast\mun\mup$ & ``Theory'' \\ \hline 
$\thetal$ & $\theta^{\rm GS}_\ell$ &  $ \pi - \theta^{\rm GS}_\ell$ \\ 
$\theta_K$ & $\theta^{\rm GS}_V$ &  $\theta^{\rm GS}_V$ \\ 
$\chi$ & $-\chi^{\rm GS}$ &  $\chi^{\rm GS}$ 
\end{tabular}
\caption{Translation dictionary between GS~\cite{Gilman:1989uy}, LHCb $\Bzb \to \overline{K}^\ast \mup\mun$~\cite{Aaij:2013iag,Aaij:2015oid} and ``theory''~\cite{Egede:2010zc} conventions for the electroweak penguin $\Bb$ decay. The latest LHCb $\Bzb \to \Km \pip \mun\mup$~\cite{lhcb_spd} paper follows the GS conventions.}
\label{tab:ewp_dict}
\end{table}

\section{The 77 spin-4 $SPDFG$-moments}

For the case where the $X_J$ system comprises spin states till spin $J=4$, there are 26 complex amplitudes $\{S^\eta, H^\eta_\lambda, D^\eta_\lambda, F^\eta_\lambda, G^\eta_\lambda\}$, where $\eta = \pm1$ tags the handedness of the leptonic side current, and $\lambda \in \{0,\pm1\}$ is the helicity of the $X_J$ system. The expressions for the helicity amplitudes in terms of the underlying QCD form-factors and Wilson coefficients can be found in Refs.~\cite{Lu:2011jm,Dey:2015rqa}. The formalism of the moments expansion as in Ref.~\cite{Dey:2015rqa} for the $J\leq2$ $SPD$-wave case is easily extended to the $J\leq 4$ $SPDFG$-wave case. From the $F$-wave onwards, every higher spin adds 18 moments and there are 77 angular moments for the spin-4 case. The differential rate is
\begin{subequations}
\label{eqn:vector_moments}
\begin{align}
\frac{d\Gamma }{d\qsq d\Omega} &= \mathcal{C} \times \left\{ \displaystyle \sum^{77}_{i=1} f_i (\Omega) \Gamma_i(\qsq) \right\} \\
\Gamma_i(\qsq) &= \Gamma^L_i(\qsq) + \eta^{L\to R}_i\; \Gamma^R_i(\qsq),
\end{align}
\end{subequations}
where the sign $\eta^{L\to R}_i=\pm 1$ depends on the signature of $f_i$ under $\thetal \to \pi + \thetal$. The orthonomal angular basis is constructed out of the the spherical harmonics $Y^m_l \equiv Y^m_l (\thetal,\chi)$ and the reduced spherical harmonics ${P^m_l \equiv \sqrt{2 \pi}Y^m_l(\thetav,0)}$. Following the notation as in Ref.~\cite{Dey:2015rqa}, the pre-factor is
\begin{equation}
\mathcal{C} = \frac{ \sqrt{8 \pi}  |V|^2 {\bf k}\; G_F^2 \qsq \mathcal{B}^{X} }{3 m^2_B (4 \pi)^4},
\end{equation}
where ${\bf k}$ is the breakup momentum if the $B$ meson and $\mathcal{B}^{X}$ is the branching fraction of the $X_J$ system into the final hadronic states uder consideration. For semileptonic $b\to q \ellm \barnuell$ decays, $V\equiv V_{bq}$, while for the electroweak transition $b\to s \ellm \ellp$,
\begin{align} 
V\equiv \displaystyle \left(\frac{\alpha}{2 \pi} V^\ast_{ts} V_{tb} \right).
\end{align}

To facilitate the discussion, we first redefine the hadronic-side amplitudes as
\begin{subequations}
\begin{align}
a &= \displaystyle \phantom{\pm} \left(2 S - \sqrt{5} D_0 + \frac{9}{4} G_0\right) \\
b &= \displaystyle \phantom{\pm} \left(2 \sqrt{3} H_0 - 3 \sqrt{7} F_0 \right) \\
c &= \displaystyle \phantom{\pm} \left(3 \sqrt{5} D_0 - \frac{45}{2} G_0 \right) \\
d &= \displaystyle \phantom{\pm} \left(5 \sqrt{7} F_0 \right) \\
e &= \displaystyle \phantom{\pm} \left(\frac{105}{4} G_0 \right) \\
l_\pm &= \mp \left(\sqrt{6} H_\pm - \frac{\sqrt{21}}{2} F_\pm  \right) \\
m_\pm &= \mp \left(\sqrt{30} D_\pm - \frac{9}{2} \sqrt{5} G_\pm \right) \\
n_\pm &= \mp \left(\frac{5 \sqrt{21}}{2} F_\pm  \right) \\
p_\pm &= \mp \left(\frac{21 \sqrt{5}}{2} G_\pm \right),
\end{align}
\label{eqn:amps_comb}
\end{subequations}
where, for sake of notational simplicity, we have removed the $\eta$ tag. For what follows, all amplitudes are understood to have an additional $\eta$ tag.

We next define the set of bilinears \{$\alpha_i$, $\beta^\pm_i$, $\delta_i$, $U_i$, $V_i$, $X_i$, $K_i$, $L_i$, $\epsilon^\pm_j$, $R_j$, $S_j$, $T_j$, $Z_j$\}, where $i \in \{0,\ldots,8\}$ and $j \in \{0,\ldots,7\}$. The expressions for these in terms of the amplitudes are tabulated in App.~\ref{sec:app_exp_bilinear_vars}. We also define a set of coefficients $e_l$ and $s_l$ to construct the 77 orthonormal basis moments in Tables~\ref{tab:spdfg_1_mom} and \ref{tab:spdfg_2_mom}. The numeric values of the $e_l$ and $s_l$ coefficents are also defined in App.~\ref{sec:app_exp_coeff_vars}.

Table~\ref{tab:spdfg_1_mom} lists the 41 moments for the $SPD$-wave case, with additional contributions from the higher waves incorporated now. Table~\ref{tab:spdfg_2_mom} lists the additional 36 moments for the $SPDFG$ case. The pattern is easily discernable now -- from the $F$-wave onwards, each additional higher wave adds 18 angular moments.

\begin{table*}
\centering
\begin{tabular}{c|c|c|c}
 $i$    &   $f_i(\Omega)$             & $\Gamma^{L, {\rm tr}}_i(\qsq)$ & $\eta^{L\to R}_i$ \\ \hline \hline
 1   &   $P^0_0 Y^0_0$     &  $\left[ \Uze + \ezo \Utw + \eztw \Uf + \ezth \Usi + \ezf \Ue \right]$  & + ($L \to R$)\\ \hline
 2   &   $P^0_1 Y^0_0$     &  $\left[ \eoz \Uo + \eoo \Uth + \eotw \Ufi + \eoth \Use \right] $  & " \\ \hline
 3   &   $P^0_2 Y^0_0$     &  $\left[ \etwz \Utw + \etwo \Uf + \etwtw \Usi + \etwth \Ue \right] $  & " \\ \hline
 4   &   $P^0_3 Y^0_0$     &  $\left[ \ethz \Uth + \etho \Ufi + \ethtw \Use \right] $  & "  \\ \hline
 5   &   $P^0_4 Y^0_0$     &  $\left[ \efz \Uf + \efo \Usi + \eftw \Ue \right] $  & "  \\ \hline
 6   &   $P^0_0 Y^0_2$     &  $\left[ \Vze + \ezo \Vtw + \eztw \Vf + \ezth \Vsi + \ezf \Ve \right] $  & " \\ \hline
 7   &   $P^0_1 Y^0_2$     &  $\left[ \eoz \Vo + \eoo \Vth + \eotw \Vfi + \eoth \Vse \right] $  & " \\ \hline
 8   &   $P^0_2 Y^0_2$     &  $\left[ \etwz \Vtw + \etwo \Vf + \etwtw \Vsi + \etwth \Ve \right] $  & "  \\ \hline
 9   &   $P^0_3 Y^0_2$     &  $\left[ \ethz \Vth + \etho \Vfi + \ethtw \Vse \right] $  & "  \\ \hline
 10  &   $P^0_4 Y^0_2$     &  $\left[ \efz \Vf + \efo \Vsi + \eftw \Ve \right] $  & "  \\ \hline
 11  &   $P^1_1 \sqrt{2}\rel(Y^1_2)$ &  $\left[ \Zz + \szo \Ztw+ \sztw \Zf + \szth \Zsi \right] $  & " \\ \hline
 12  &   $P^1_2 \sqrt{2}\rel(Y^1_2)$ &  $\left[ \soz \Zo + \soo \Zth+ \sotw \Zfi + \soth \Zse \right] $  & " \\ \hline
 13  &   $P^1_3 \sqrt{2}\rel(Y^1_2)$ &  $\left[ \stwz \Ztw + \stwo \Zf + \stwtw \Zsi \right] $  & "  \\ \hline
 14  &   $P^1_4 \sqrt{2}\rel(Y^1_2)$ &  $\left[ \sthz \Zth + \stho \Zfi + \sthtw \Zse\right] $  & "  \\ \hline
 15  &   $P^1_1 \sqrt{2}\img(Y^1_2)$ &  $\left[ \Sz + \szo \Stw+ \sztw \Sf + \szth \Ssi \right] $  & "  \\ \hline
 16  &   $P^1_2 \sqrt{2}\img(Y^1_2)$ &  $\left[ \soz \So + \soo \Sth+ \sotw \Sfi + \soth \Sse \right] $  & " \\ \hline
 17  &   $P^1_3 \sqrt{2}\img(Y^1_2)$ &  $\left[ \stwz \Stw + \stwo \Sf + \stwtw \Ssi \right] $  & "  \\ \hline
 18  &   $P^1_4 \sqrt{2}\img(Y^1_2)$ &  $\left[ \sthz \Sth + \stho \Sfi + \sthtw \Sse \right] $  & " \\ \hline
 19  &   $P^0_0 \sqrt{2}\rel(Y^2_2)$ &  $\left[\Kze + \ezo \Ktw + \eztw \Kf + \ezth \Ksi + \ezf \Ke \right] $  & " \\ \hline
 20  &   $P^0_1 \sqrt{2}\rel(Y^2_2)$ &  $\left[\eoz \Ko + \eoo \Kth + \eotw \Kfi + \eoth \Kse \right] $  & " \\ \hline
 21  &   $P^0_2 \sqrt{2}\rel(Y^2_2)$ &  $\left[\etwz \Ktw + \etwo \Kf + \etwtw \Ksi + \etwth \Ke \right] $  & " \\ \hline
 22  &   $P^0_3 \sqrt{2}\rel(Y^2_2)$ &  $\left[\ethz \Kth + \etho \Kfi + \ethtw \Kse \right] $  & " \\ \hline
 23  &   $P^0_4 \sqrt{2}\rel(Y^2_2)$ &  $\left[\efz \Kf + \efo \Ksi + \eftw \Ke \right] $  & " \\ \hline
 24  &   $P^0_0 \sqrt{2}\img(Y^2_2)$ &  $\left[\Lze + \ezo \Ltw + \eztw \Lf + \ezth \Lsi + \ezf \Le \right] $  & " \\ \hline
 25  &   $P^0_1 \sqrt{2}\img(Y^2_2)$ &  $\left[\eoz \Lo + \eoo \Lth + \eotw \Lfi + \eoth \Lse \right] $  & " \\ \hline
 26  &   $P^0_2 \sqrt{2}\img(Y^2_2)$ &  $\left[\etwz \Ltw + \etwo \Lf + \etwtw \Lsi + \etwth \Le \right] $  & " \\ \hline
 27  &   $P^0_3 \sqrt{2}\img(Y^2_2)$ &  $\left[\ethz \Lth + \etho \Lfi + \ethtw \Lse \right] $  & " \\ \hline
 28  &   $P^0_4 \sqrt{2}\img(Y^2_2)$ &  $\left[\efz \Lf + \efo \Lsi + \eftw \Le \right] $  & " \\ \hline 
 29  &   $P^0_0 Y^0_1$ &  $\left[\Xze + \ezo \Xtw + \eztw \Xf + \ezth \Xsi + \ezf \Xe \right] $  & - ($L \to R$) \\ \hline
 30  &   $P^0_1 Y^0_1$ &  $\left[\eoz \Xo + \eoo \Xth + \eotw \Xfi + \eoth \Xse \right] $  & " \\ \hline
 31  &   $P^0_2 Y^0_1$ &  $\left[\etwz \Xtw + \etwo \Xf + \etwtw \Xsi + \etwth \Xe \right] $  & " \\ \hline
 32  &   $P^0_3 Y^0_1$ &  $\left[\ethz \Xth + \etho \Xfi + \ethtw \Xse \right] $  & " \\ \hline
 33  &   $P^0_4 Y^0_1$ &  $\left[\efz \Xf + \efo \Xsi + \eftw \Xe \right] $  & " \\ \hline 
 34  &   $P^1_1 \sqrt{2}\rel(Y^1_1)$ &  $\left[ \Rz + \szo \Rtw+ \sztw \Rf + \szth \Rsi \right] $  & "  \\ \hline
 35  &   $P^1_2 \sqrt{2}\rel(Y^1_1)$ &  $\left[ \soz \Ro + \soo \Rth+ \sotw \Rfi + \soth \Rse \right] $  & " \\ \hline
 36  &   $P^1_3 \sqrt{2}\rel(Y^1_1)$ &  $\left[ \stwz \Rtw + \stwo \Rf + \stwtw \Rsi \right] $  & "  \\ \hline
 37  &   $P^1_4 \sqrt{2}\rel(Y^1_1)$ &  $\left[ \sthz \Rth + \stho \Rfi + \sthtw \Rse \right] $  & " \\ \hline
 38  &   $P^1_1 \sqrt{2}\img(Y^1_1)$ &  $\left[ \Tz + \szo \Ttw+ \sztw \Tf + \szth \Tsi \right] $  & "  \\ \hline
 39  &   $P^1_2 \sqrt{2}\img(Y^1_1)$ &  $\left[ \soz \To + \soo \Tth+ \sotw \Tfi + \soth \Tse \right] $  & " \\ \hline
 40  &   $P^1_3 \sqrt{2}\img(Y^1_1)$ &  $\left[ \stwz \Ttw + \stwo \Tf + \stwtw \Tsi \right] $  & "  \\ \hline
 41  &   $P^1_4 \sqrt{2}\img(Y^1_1)$ &  $\left[ \sthz \Tth + \stho \Tfi + \sthtw \Tse \right] $  & " \\ \hline
\end{tabular}
\caption{$SPDFG$-wave moments: the first 41 angular moments corresponding to those for the $SPD$-only case, extended to include $F$- and $G$-wave contributions.}
\label{tab:spdfg_1_mom}
\end{table*}

\begin{table}
\centering
\begin{tabular}{c|c|c|c}
 $i$    &   $f_i(\Omega)$             & $\Gamma^{L, {\rm tr}}_i(\qsq)$ & $\eta^{L\to R}_i$ \\ \hline \hline
  \multicolumn{4}{c}{addition of $F$-wave} \\ \hline
  42  &   $P^0_5 Y^0_0$     &  $\left[ \efiz \Ufi + \efio \Use \right]$ & + ($L \to R$) \\ \hline
  43  &   $P^0_6 Y^0_0$     &  $\left[ \esiz \Usi + \esio \Ue \right]$  & " \\ \hline
  44  &   $P^0_5 Y^0_2$     &  $\left[ \efiz \Vfi + \efio \Vse \right]$  & " \\ \hline
  45  &   $P^0_6 Y^0_2$     &  $\left[ \esiz \Vsi + \esio \Ve  \right]$  & "\\ \hline
  46  &   $P^1_5 \sqrt{2}\rel(Y^1_2)$ &  $\left[\sfz \Zf + \sfo \Zsi \right] $  & " \\ \hline
  47  &   $P^1_6 \sqrt{2}\rel(Y^1_2)$ &  $\left[\sfiz \Zfi + \sfio \Zse \right] $  & " \\ \hline
  48  &   $P^1_5 \sqrt{2}\img(Y^1_2)$ &  $\left[\sfz \Sf + \sfo \Ssi \right] $  & " \\ \hline
  49  &   $P^1_6 \sqrt{2}\img(Y^1_2)$ &  $\left[\sfiz \Sfi + \sfio \Sse \right] $  & " \\ \hline
  50  &   $P^0_5 \sqrt{2}\rel(Y^2_2)$ &  $\left[\efiz \Kfi + \efio \Kse \right] $  & "\\ \hline
  51  &   $P^0_6 \sqrt{2}\rel(Y^2_2)$ &  $\left[\esiz\Ksi + \esio \Ke \right] $  & "\\ \hline
  52  &   $P^0_5 \sqrt{2}\img(Y^2_2)$ &  $\left[\efiz \Lfi + \efio \Lse \right] $  & "\\ \hline
  53  &   $P^0_6 \sqrt{2}\img(Y^2_2)$ &  $\left[\esiz\Lsi + \esio \Le \right] $  & "\\ \hline
  54  &   $P^0_5 Y^0_1$     &  $\left[ \efiz \Xfi + \efio \Xse \right]$ & - ($L \to R$) \\ \hline
  55  &   $P^0_6 Y^0_1$     &  $\left[ \esiz \Xsi + \esio \Xe \right]$  & " \\ \hline
  56  &   $P^1_5 \sqrt{2}\rel(Y^1_1)$ &  $\left[\sfz \Rf + \sfo \Rsi \right] $  & " \\ \hline
  57  &   $P^1_6 \sqrt{2}\rel(Y^1_1)$ &  $\left[\sfiz \Rfi + \sfio \Rse \right] $  & " \\ \hline
  58  &   $P^1_5 \sqrt{2}\img(Y^1_1)$ &  $\left[\sfz \Tf + \sfo \Tsi \right] $  & " \\ \hline
  59  &   $P^1_6 \sqrt{2}\img(Y^1_1)$ &  $\left[\sfiz \Tfi + \sfio \Tse \right] $  & " \\ \hline
  \multicolumn{4}{c}{addition of $G$-wave} \\ \hline
  60  &   $P^0_7 Y^0_0$     &  $ \ese \Use $ & + ($L \to R$) \\ \hline
  61  &   $P^0_8 Y^0_0$     &  $ \ee \Ue $  & "\\ \hline
  62  &   $P^0_7 Y^0_2$     &  $ \ese \Vse $  & "\\ \hline
  63  &   $P^0_8 Y^0_2$     &  $ \ee \Ve $  & " \\ \hline
  64  &   $P^1_7 \sqrt{2}\rel(Y^1_2)$ &  $ \ssi \Zsi $  & "\\ \hline
  65  &   $P^1_8 \sqrt{2}\rel(Y^1_2)$ &  $ \sse \Zse $  & "\\ \hline
  66  &   $P^1_7 \sqrt{2}\img(Y^1_2)$ &  $ \ese \Lse $  & "\\ \hline
  67  &   $P^1_8 \sqrt{2}\img(Y^1_2)$ &  $ \ee \Le  $  & "\\ \hline
  68  &   $P^0_7 \sqrt{2}\rel(Y^2_2)$ &  $ \ese \Kse $  & "\\ \hline
  69  &   $P^0_8 \sqrt{2}\rel(Y^2_2)$ &  $ \ee \Ke $  & "\\ \hline
  70  &   $P^0_7 \sqrt{2}\img(Y^2_2)$ &  $ \ese \Lse  $  & "\\ \hline
  71  &   $P^0_8 \sqrt{2}\img(Y^2_2)$ &  $ \ee \Le $  & "\\ \hline
  72  &   $P^0_7 Y^0_1$     &  $ \ese \Xse $ & - ($L \to R$) \\ \hline
  73  &   $P^0_8 Y^0_1$     &  $ \ee \Xe $  & "\\ \hline
  74  &   $P^1_7 \sqrt{2}\rel(Y^1_1)$ &  $ \ssi \Rsi $  & "\\ \hline
  75  &   $P^1_8 \sqrt{2}\rel(Y^1_1)$ &  $ \sse \Rse $  & "\\ \hline
  76  &   $P^1_7 \sqrt{2}\img(Y^1_1)$ &  $ \ssi \Tsi $  & "\\ \hline
  77  &   $P^1_8 \sqrt{2}\img(Y^1_1)$ &  $ \sse \Tse $  & "\\ \hline
\end{tabular}
\caption{$SPDFG$-wave moments: the 18 + 18 set of additional moments over those in Table~\ref{tab:spdfg_1_mom} for the $SPD$ case.}
\label{tab:spdfg_2_mom}
\end{table}

\section{Conistency relations for $SPD$-waves}

\subsection{The two-component notation}

To facilitate the discussion, extending the notation developed in Ref.~\cite{Hofer:2015kka}, we define the two-component complex vectors for $S$-, $P$- and $D$-waves for electroweak penguins:


\begin{alignat}{5}
 \label{eqn:bilinears}
 s&= \left( \begin{array}{c} S^L \\S^{R \ast} \end{array} \right) & & & & \nonumber \\ 
 h_\parallel &= \left( \begin{array}{c} H^L_\parallel \\H^{R \ast}_\parallel \end{array} \right) & 
 \;\;h_\perp &= \left( \begin{array}{c} H^L_\perp\\ - H^{R \ast}_\perp \end{array} \right) & \;\;h_0 &= \left( \begin{array}{c} H^L_0\\ H^{R \ast}_0 \end{array} \right) \nonumber \\
 d_\parallel &= \left( \begin{array}{c} D^L_\parallel \\D^{R \ast}_\parallel \end{array} \right) & 
 \;\;d_\perp &= \left( \begin{array}{c} D^L_\perp\\ - D^{R \ast}_\perp \end{array} \right) & \;\;d_0 &= \left( \begin{array}{c} D^L_0\\ D^{R \ast}_0 \end{array} \right).
 \end{alignat}
The $L(R)$ superscripts denote the left(LH)- and right-handed(RH) amplitudes, respectively, and this structure can be continued further to any spin-$J$. The physical observables will then be constructed out of bilinears formed out of these two component objects.

\subsection{The two-fold ambiguity and the 17 consistency relations}
\label{sec:ambiguities}

For the charged di-lepton cases, since the final leptonic spins are not measured but averaged over, the full rate is invariant under the following global transformation of each spin-$J$ helicity amplitude:
\begin{align}
\label{eqn:twofoldamb}
\mathcal{H}^{\eta,J}_\lambda \to \left(\mathcal{H}^{-\eta,J}_{-\lambda}\right)^\ast.
\end{align}
For the electromagnetic $c \bar{c} \to \ell^+ \ell^-$ decays, the LH and RH amplitudes are equal and Eq.~\ref{eqn:twofoldamb} represents the same two-fold ambiguity as in the determination of $\beta$ and $\beta_s$ from $B \to J/\psi K\pi$~\cite{Aubert:2004cp} and  $B_s \to J/\psi KK$~\cite{Aaij:2014zsa}, respectively. 

The effect of this transformation on the two-component vectors in Eq.~\ref{eqn:bilinears} is that the top and bottom rows get swapped. Therefore, for any pair $\{n_i,n_j\}$ of the two-component vectors, the observable $\mathcal{O}_{ij}\equiv n^\dagger_i n_j$ is invariant under $n_i\to U n_i$, where the group of symmetry transformation is the group $U(2)$ with $n_{\rm gen} = 4$ generators and all physical observables allowed in the angular rate expression is of the type $\mathcal{O}_{ij}$.

For the $SPD$-wave case, with 14 amplitudes, there are $2 n_A=28$ independent real variables and $n_{\rm obs} = 2 n_A - n_{\rm gen} = 24$ real observables. However, the number of angular moments from Ref.~\cite{Dey:2015rqa} is $n_{\rm mom} = 41$. Hence, there must be $n_{\rm rel} = n_{\rm mom} - n_{\rm obs} = 17$ relations amongst the 41 observables.

\subsection{The double-basis ``trick''}
\label{sec:double_basis}

The key towards finding these additional relations is the fact that of the seven $n_i\in \{h_\parallel,h_\perp,h_0,d_\parallel,d_\perp,d_0,s\}$ in Eq.~\ref{eqn:bilinears}, any two can be chosen as a basis set and the rest can expressed as a linear combination. Denoting the basis as $\{n_1,n_2\}$, any other $n_i$ can be written as
\begin{align}
n_i &= a_i n_1 + b_i n_2,
\end{align}
where the coefficients $a_i$, $b_i$ can be solved as~\cite{Hofer:2015kka}
\begin{align}
a_i & = \displaystyle \frac{|n_2|^2 (n^\dagger_1 n_i) - (n^\dagger_1 n_2)  (n^\dagger_2 n_i)}{|n_1|^2 |n_2|^2 - |n^\dagger_1 n_2|^2}\\
b_i & = \displaystyle \frac{|n_1|^2 (n^\dagger_2 n_i) - (n^\dagger_2 n_1)  (n^\dagger_1 n_i)}{|n_1|^2 |n_2|^2 - |n^\dagger_1 n_2|^2}
\end{align}
Thus we have 5 real equations:
\begin{align}
|n_i|^2 &= a_i (n^\dagger_i n_1) + b_i (n^\dagger_i n_2), \;\; i\in \{3,...,7\}
\end{align}
and $^5C_2 = 10$ complex equations:
\begin{align}
n^\dagger_i n_j & = a_j(n^\dagger_i n_1) + b_j(n^\dagger_i n_2),\;\; i,j\in \{3,...,7\}
\end{align}
If one had expressions of all the relevant $Re(n_i^\dagger n_j)$ and $Im(n_i^\dagger n_j)$ in terms of the observables, the above sets of equations would yield all the relations. 

For the pure $P$-wave case, the number of independent observables is 8 and all the combinations $n^\dagger_i n_j$ can be solved out in terms of the observables~\cite{Hofer:2015kka}. For the $SP$-wave case, there are 12 independent onservables. Even here, the combination $Im(n^\dagger_S n_0)$ is not given in terms of any of the angular observables. It is fortuitous that this combination is not needed to solve for the relations. The minimal set of the 12 independent observables for the $SP$-wave case was given in Ref.~\cite{Hofer:2015kka}.

For the $SPD$-wave case, the situation is completely different, and many of the $n^\dagger_i n_j$ combinations remain unsolvable in terms of the observables.

\subsection{Solving for the bilinears}
\label{sec:bil_solns}

For the $SPD$-wave case, we list below the known set of bilinears that can be expressed in terms of the 41 $\Gamma_i$ moments observables in Ref.~\cite{Dey:2015rqa}:
\begin{subequations}
\renewcommand{\theequation}{\theparentequation.\arabic{equation}}
\begin{align}
\qdzsq &= \frac{7}{9} \left( \frac{\Gamma_5}{2} - \sqrt{5} \Gamma_{10} \right) \\
\qdpasq &= \frac{7}{4} \left( \sqrt{\frac{5}{3}} \Gamma_{23}  - \frac{1}{3} \left( \sqrt{5} \Gamma_{10} + \Gamma_5    \right)   \right)\\
\qdpesq &= \frac{7}{4} \left( -\sqrt{\frac{5}{3}} \Gamma_{23}  - \frac{1}{3} \left( \sqrt{5} \Gamma_{10} + \Gamma_5    \right)   \right)\\
\qhpasq &= \frac{1}{2} \Bigg[ \frac{2}{3} \left(\Gamma_1 + \sqrt{5} \Gamma_6 \right) + \frac{7}{6} \left( \sqrt{5} \Gamma_{10} + \Gamma_5 \right) \nonumber \\
 & \hspace{0.7cm} - \sqrt{ \frac{5}{3} } \left(2 \Gamma_{19} + \frac{7}{2} \Gamma_{23} \right) \Bigg] \\
\qhpesq &= \frac{1}{2} \Bigg[ \frac{2}{3} \left(\Gamma_1 + \sqrt{5} \Gamma_6 \right) + \frac{7}{6} \left( \sqrt{5} \Gamma_{10} + \Gamma_5 \right) \nonumber \\
 & \hspace{0.7cm} + \sqrt{ \frac{5}{3} } \left(2 \Gamma_{19} + \frac{7}{2} \Gamma_{23} \right) \Bigg]\\
\qrhpadpa &= \frac{5}{6} \left[ \frac{1}{\sqrt{3}} \left(\frac{\Gamma_2}{\sqrt{5}} + \Gamma_7\right) -\Gamma_{20} \right]\\
\qrhpedpe &= \frac{5}{6} \left[ \frac{1}{\sqrt{3}} \left(\frac{\Gamma_2}{\sqrt{5}} + \Gamma_7\right) +\Gamma_{20} \right]\\
\qrhzdz   &= \frac{1}{3\sqrt{3}} \left[ \frac{\sqrt{35}}{2} \Gamma_4 + \frac{5}{\sqrt{3}} \left( \Gamma_7 + \frac{\Gamma_2}{\sqrt{5}} \right) \right]\\
\qrshz &= - \frac{1}{18} \Gamma_2 - \frac{\sqrt{21}}{9} \Gamma_4 - \frac{5 \sqrt{5}}{9} \Gamma_7 \\
\qrhpehpa &= \frac{5}{12} \left( - \frac{\Gamma_{29}}{\sqrt{3}} + \frac{7}{\sqrt{15}} \Gamma_{31} \right)\\ 
\qihpehpa &= \frac{5}{12 \sqrt{3}} \left( \sqrt{5} \Gamma_{24} - 7 \Gamma_{26} \right) \\
\qrdpedpa &= -\frac{7}{12\sqrt{3}} \left( \Gamma_{29} + \sqrt{5} \Gamma_{31} \right) \\
\qidpedpa &= \frac{7}{12}  \sqrt{\frac{5}{3}} \left( \Gamma_{24} + \sqrt{5} \Gamma_{26} \right) \\
\qrdpadz &= - \frac{7 \sqrt{2}}{6} \Gamma_{14} \\
\qidpadz &=  - \frac{7}{3 \sqrt{10}} \Gamma_{41} \\
\qrdpedz &=  \frac{7}{3 \sqrt{10}} \Gamma_{37} \\
\qidpedz &=  \frac{7\sqrt{2}}{6} \Gamma_{18} 
\end{align}
\end{subequations}

In particular $(\qhzsq + \sqrt{5} \qrsdz)$ occurs together, so $\qhzsq$ can not be extracted. Also, $(\qhzsq + \qssq)$ occurs together, so $F_S$ can not be extracted. The $D$-wave fraction
\begin{align}
F_D &= (\qdzsq + \qdpesq + \qdpasq)/\Gamma_1 
\end{align}
is completely extractable however. 

\subsection{Relations among the observables}

\begin{figure}
\centering
\includegraphics[width=0.45\textwidth]{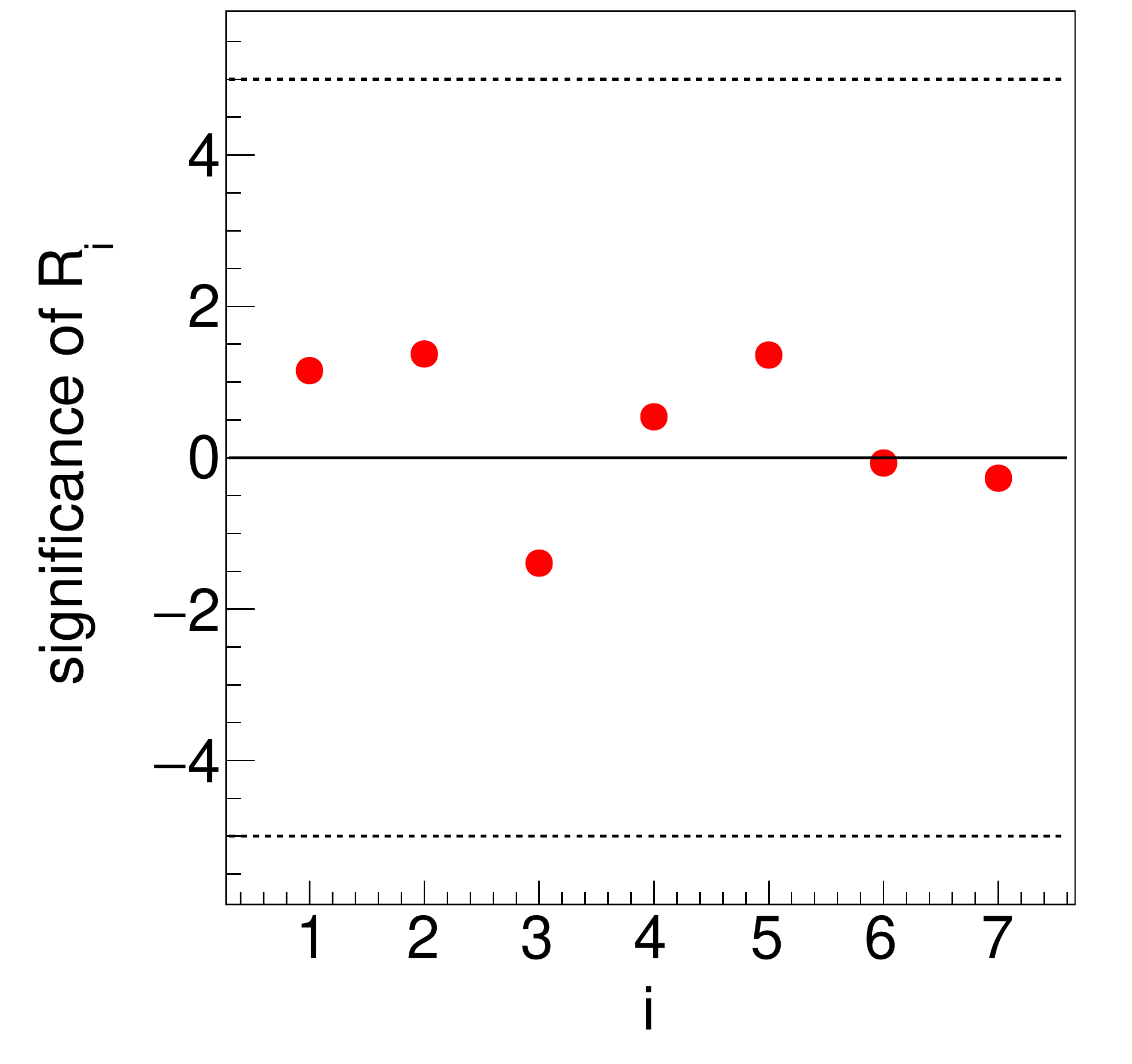}
\caption{The significance of the seven $R_i$ observables defined in Eq.~\ref{eqn:linear_consistency_rel}, integrated over the range $\qsq \in [1.1,6.0]$~GeV$^2$ and $\mkpi \in [1330,1530]$~MeV, as measured by LHCb in the mode $\Bzb \to \Km \pip \mun \mup$~\cite{lhcb_spd}. No significant deviation from zero is seen, pointing to the internal consistency among the measurements.} 
\label{fig:consistency_rel}
\end{figure}

We first list the 7 linear relations below:
\begin{subequations}
\label{eqn:linear_consistency_rel}
\renewcommand{\theequation}{\theparentequation.\arabic{equation}}
\begin{align}
0 = R_1 &\equiv \Gamma_{25} + \sqrt{\frac{7}{3}} \Gamma_{27} \\
0 = R_2 &\equiv \Gamma_{20} + \sqrt{\frac{7}{3}} \Gamma_{22} \\
0 = R_3 &\equiv  \sqrt{\frac{3}{7}} \Gamma_{30} + \Gamma_{32} \\
0 = R_4 &\equiv \frac{1}{\sqrt{5}} \left( 3 \Gamma_{23} + \Gamma_{19} \right) + \Gamma_{21}\\
0 = R_5 &\equiv \frac{\Gamma_3}{\sqrt{5}} +  \Gamma_8 \nonumber \\
        & \;\; + \frac{3}{5} \left[ (\sqrt{5} \Gamma_{10} + \Gamma_5) + \frac{1}{3} (\Gamma_1 + \sqrt{5} \Gamma_6) \right]\\
0 = R_6 &\equiv 3 \Gamma_{28} +  \left( \Gamma_{24} + \sqrt{5} \Gamma_{26} \right)\\
0 = R_7 &\equiv 3 \Gamma_{33} + \left( \Gamma_{29} + \sqrt{5} \Gamma_{31} \right),
\end{align}
\end{subequations}
where the $R_i$'s can be taken as new observables as functions of $\qsq$ and $m_X$ whose measured values are expected to be zero. A useful consistency check is therefore that the experimental data show no significant deviation from zero in these observables. Figure~\ref{fig:consistency_rel} shows the experimentally measured significance (measured value divided by its uncertainty) from LHCb~\cite{lhcb_spd} of the seven $R_i$ observables integrated over the $\qsq$ range $[1.1,6.0]$~GeV$^2$ and \mkpi range $[1330,1530]$~MeV in the mode $\Bzb \to \Km \pip \mun \mup$. All seven $R_i$ observables are found consistent with zero.

Next, we employ the method elaborated in Sec.~\ref{sec:double_basis} using the basis $\{d_0,d_\perp\}$ and $n_i = d_\parallel$. This yields the following relation: 
\begin{widetext}
\begin{align}
0 = R_8 &\equiv \left[ \left( \sqrt{ \frac{5}{3}} \Gamma_{23} + \frac{  \sqrt{5} \Gamma_{10} + \Gamma_5 }{3} \right) \left( \Gamma_{14}^2 + \frac{\Gamma_{41}^2}{5} \right) - \left(\frac{\Gamma_5/2 - \sqrt{5} \Gamma_{10}  }{54}\right) \left(  (\Gamma_{29} + \sqrt{5} \Gamma_{31})^2 + 5 (\Gamma_{24} + \sqrt{5} \Gamma_{26})^2 \right)   \right] \nonumber \\
  &  \;\;\; +\frac{2}{3 \sqrt{15}} \left[\left(\Gamma_{37} \Gamma_{14} + \Gamma_{18} \Gamma_{41}\right) \left(\Gamma_{29} + \sqrt{5} \Gamma_{31}\right) + \left( \Gamma_{37} \Gamma_{41} - 5 \Gamma_{18} \Gamma_{14}\right) \left(\Gamma_{24} + \sqrt{5} \Gamma_{26}\right) \right] \nonumber \\
 & \;\;\; - \left( \sqrt{ \frac{5}{3}} \Gamma_{23} - \frac{  \sqrt{5} \Gamma_{10} + \Gamma_5 }{3} \right) \left[ \left( \frac{\Gamma_5/2 - \sqrt{5} \Gamma_{10}}{2} \right) \left( \sqrt{ \frac{5}{3}} \Gamma_{23} + \frac{  \sqrt{5} \Gamma_{10} + \Gamma_5 }{3} \right)  + \left( \frac{\Gamma_{37}^2}{5} + \Gamma_{18}^2 \right) \right]
\end{align}
\end{widetext}
Since this relation is non-linear, it is not possible to validate this from experimentally measured binned observables $\Gamma_i$ that are integrated over a given $\qsq$ and $m_X$ range. Within the triplet, $\{d_0, d_\perp, d_\parallel \}$, we have also checked that all other combinations yield no new relation.

\section{Conclusions}
\label{sec:conc}

In conclusion, we have presented the the first full angular moments expansion for semileptonic and electroweak penguin $\Bb\to X_J \ell_1 \ell_2$ decays, where the $X_J$ system can be in a spin $J\leq4$ state. This will enable angular analyses of even higher spin states towards New Physics searches incorporating these new exited modes. We expect this method to be directly employed in Run~II analyses at LHCb. Further applications can include $\Bbar \to \pi \pi \ellm \barnuell$ over a wide $m_{\pi\pi}$ kinematic regime at Belle~II, for extracting $\Vub$ and right-handed current searches. The method can also be used to directly probe the spin content of excited $D^{\ast \ast}$ states in exclusive $\Bbar \to D^{\ast \ast} \ellm \barnuell$ topologies towards reconciling the existing gap between inclusive and exclusive $\Bbar \to X_c \ellm \barnuell$ branching fractions~\cite{Bernlochner:2012bc}.

The formalism also allows for internal consistency checks, since the moments are not independent observables. In this paper we also validated the available consistency checks for the latest LHCb analysis~\cite{lhcb_spd} involving $S$-, $P$- and $D$-waves.

\begin{acknowledgements}
We thank Quim Matias for helpful discussions on the two-component notation and derivations of the consistency relations. 
\end{acknowledgements}

\appendix
\section{Bilinear variables for the expansion}
\label{sec:app_exp_bilinear_vars}

We list here the definitions of the various bilinears in terms of the amplitudes given by Eq.~\ref{eqn:amps_comb}.

\begin{subequations}
\begin{flalign}
\alpha_0 &= \left(|a|^2 \right) & \\
\alpha_1 &= \left(a b^\ast + b a^\ast \right) \\
\alpha_2 &= \left(a c^\ast + c a^\ast + |b|^2 \right) \\
\alpha_3 &= \left(a d^\ast + d a^\ast + b c^\ast + c b^\ast \right) \\
\alpha_4 &= \left(a e^\ast + a^\ast e + b d^\ast + d^\ast b + |c|^2 \right) \\
\alpha_5 &= \left(b e^\ast + e b^\ast + c d^\ast + d c^\ast \right) \\
\alpha_6 &= \left(c e^\ast + e c^\ast + |d|^2 \right) \\
\alpha_7 &= \left(e d^\ast + e^\ast d \right) \\
\alpha_8 &= \left(|e|^2 \right)
\end{flalign}
\end{subequations}

\begin{subequations}
\begin{flalign}
\delta_0 &= \left(\lp\lms \right) & \\
\delta_1 &= \left(\lp\mms + \mpee \lms \right) \\
\delta_2 &= \left(\lp \nms + \np\lms + \mpee \mms - \lp \lms \right) \\
\delta_3 &= \left(\lp \pms + \pp \lms + \mpee \nms + \np \mms - (\lp\mms + \mpee \lms) \right) \\
\delta_4 &= \left(\mpee \pms + \pp \mms + \np \nms - (\lp \nms + \np \lms + \mpee \mms) \right) \\
\delta_5 &= \left(\np \pms + \pp \nms - (\lp \pms + \pp \lms + \mpee \nms + \np \mms) \right) \\
\delta_6 &= \left(\pp \pms - (\mpee \pms + \pp \mms + \np \nms)  \right) \\
\delta_7 &= -\left( \np \pms + \pp \nms\right) \\
\delta_8 &=  -\left( \pp \pms\right)
\end{flalign}
\end{subequations}

\begin{subequations}
\begin{flalign}
\beta^\pm_0 &= \left(|\lpm|^2 \right) &\\
\beta^\pm_1 &= \left(\lpm \mpms + \lpms \mpm \right) \\
\beta^\pm_2 &= \left(\lpm \npms + \lpms \npm + |\mpm|^2 - |\lpm|^2 \right) \\
\beta^\pm_3 &= \left(\lpm \ppms + \lpms \ppm + \mpm \npms + \mpms \npm - (\lpm \mpms + \lpms \mpm) \right) \\
\beta^\pm_4 &= \left(\mpm \ppms + \mpms \ppm + |\npm|^2 - (\lpm \npms + \lpms \npm + |\mpm|^2) \right) \\
\beta^\pm_5 &= \left(\npm \ppms + \npms \ppm - (\lpm \ppms + \lpms \ppm + \mpm \npms + \mpms \npm) \right) \\
\beta^\pm_6 &= \left(|\ppm|^2 - (\mpm \ppms + \mpms \ppm + |\npm|^2) \right) \\
\beta^\pm_7 &= -\left( \npm \ppms + \npms \ppm\right) \\
\beta^\pm_8 &= -\left( |\ppm|^2\right) 
\end{flalign}
\end{subequations}

\begin{subequations}
\begin{flalign}
\epsilon^\pm_0 &= \left(\lpm a^\ast \right) &\\
\epsilon^\pm_1 &= \left(\lpm b^\ast + \mpm a^\ast \right) \\
\epsilon^\pm_2 &= \left(\lpm c^\ast + \npm a^\ast + \mpm b^\ast \right) \\
\epsilon^\pm_3 &= \left(\lpm d^\ast + \ppm a^\ast + \mpm c^\ast + \npm b^\ast \right) \\
\epsilon^\pm_4 &= \left(\mpm d^\ast + \ppm b^\ast + \lpm e^\ast + \npm c^\ast  \right) \\
\epsilon^\pm_5 &= \left(\mpm e^\ast + \npm d^\ast + \ppm c^\ast \right) \\
\epsilon^\pm_6 &= \left( \npm e^\ast + \ppm d^\ast\right) \\
\epsilon^\pm_7 &= \left(\ppm e^\ast \right). 
\end{flalign}
\end{subequations}

These are used to further define the following combinations
\begin{subequations}
\begin{align}
U_i &= \phantom{-} \displaystyle \frac{\alpha_i + \beta^+_i + \beta^-_i}{4}\\
V_i &= \phantom{-} \displaystyle \frac{\beta^+_i + \beta^-_i - 2 \alpha_i}{8 \sqrt{5}}\\
X_i &= \displaystyle -\frac{\sqrt{3}}{8} (\beta^+_i - \beta^-_i) \\
K_i &= \phantom{-} \displaystyle \frac{1}{4} \sqrt{\frac{3}{5}} \rel(\delta_i) \\
L_i &= -\displaystyle \frac{1}{4} \sqrt{\frac{3}{5}} \img(\delta_i) \\
R_j &= - \displaystyle \frac{ \rel(\epsilon^+_j + \epsilon^-_j )}{4}\\
S_j &= - \displaystyle \frac{ \img(\epsilon^+_j + \epsilon^-_j )}{4 \sqrt{5}}\\
T_j &=  \phantom{-} \displaystyle \frac{ \img(\epsilon^+_j - \epsilon^-_j )}{4}\\
Z_j &=  \phantom{-} \displaystyle \frac{ \rel(\epsilon^+_j - \epsilon^-_j )}{4 \sqrt{5}},
\end{align}
\end{subequations}
for $i\in\{0,\ldots,8\}$ and $j\in\{0,\ldots,7\}$, that are employed in the definition of the moments in Tables~\ref{tab:spdfg_1_mom} and~\ref{tab:spdfg_2_mom}.

\section{The numeric coefficients}
\label{sec:app_exp_coeff_vars}

Next, we define the coefficients $e_l$ and $s_l$:

\begin{subequations}
\begin{align}
\ezo &= \displaystyle \frac{1}{3}  &  \szo &= \displaystyle \frac{1}{5}  \\
\eztw &= \displaystyle \frac{1}{5}  & \sztw &= \displaystyle \frac{3}{35}   \\
\ezth &= \displaystyle \frac{1}{7}  & \szth &= \displaystyle \frac{1}{21}   \\
\ezf &= \displaystyle \frac{1}{9}  &    \\
\eoz &= \displaystyle \frac{1}{\sqrt{3}}  & \soz &= \displaystyle \frac{1}{\sqrt{5}}   \\
\eoo &= \displaystyle \frac{\sqrt{3}}{5}  &  \soo &= \displaystyle \frac{3}{7\sqrt{5}}  \\ 
\eotw &= \displaystyle \frac{\sqrt{3}}{7}  &  \sotw &= \displaystyle \frac{\sqrt{5}}{21}  \\
\eoth &= \displaystyle \frac{1}{3\sqrt{3}}  &  \soth &= \displaystyle \frac{\sqrt{5}}{33}  \\
\etwz &= \displaystyle \frac{2}{3\sqrt{5}}  & \stwz &= \displaystyle \frac{2\sqrt{2}}{5\sqrt{7}}   \\
\etwo &= \displaystyle \frac{4}{7\sqrt{5}}  &  \stwo &= \displaystyle \frac{4\sqrt{2}}{15\sqrt{7}}  \\
\etwtw &= \displaystyle \frac{10}{21\sqrt{5}}  & \stwtw &= \displaystyle \frac{2\sqrt{2}}{11\sqrt{7}}   \\
\etwth &= \displaystyle \frac{40}{99\sqrt{5}}  &    \\
\ethz &= \displaystyle \frac{2}{5\sqrt{7}}  &  \sthz &= \displaystyle \frac{2\sqrt{2}}{7\sqrt{15}}  \\
\etho &= \displaystyle \frac{4}{9\sqrt{7}}  &  \stho &= \displaystyle \frac{4\sqrt{10}}{77\sqrt{3}}  \\
\ethtw &= \displaystyle \frac{2\sqrt{7}}{33}  &  \sthtw &= \displaystyle \frac{2\sqrt{30}}{143}  \\
\efz &= \displaystyle \frac{8}{105}  &  \sfz &= \displaystyle \frac{8}{21\sqrt{55}}  \\
\efo &= \displaystyle \frac{8}{77}  &  \sfo &= \displaystyle \frac{8\sqrt{5}}{91\sqrt{11}}  \\
\eftw &= \displaystyle \frac{16}{143}  &    \\ 
\efiz &= \displaystyle \frac{8}{63\sqrt{11}}  & \sfiz &= \displaystyle \frac{8}{33\sqrt{91}}   \\ 
\efio &= \displaystyle \frac{8}{39\sqrt{11}} &  \sfio &= \displaystyle \frac{8\sqrt{7}}{165\sqrt{13}}  \\
\esiz &= \displaystyle \frac{16}{231\sqrt{13}}  & \ssi &= \displaystyle \frac{32}{429\sqrt{35}}   \\ 
\esio &= \displaystyle \frac{64}{495\sqrt{13}} &  \sse &= \displaystyle \frac{32}{715\sqrt{51}}  \\
\ese &= \displaystyle \frac{16}{429\sqrt{15}}  &    \\
\ee &= \displaystyle \frac{128}{6435\sqrt{17}} 
\end{align}
\end{subequations}

\bibliographystyle{apsrev}
\bibliography{biblio}

\begin{thebibliography}{22}
\expandafter\ifx\csname natexlab\endcsname\relax\def\natexlab#1{#1}\fi
\expandafter\ifx\csname bibnamefont\endcsname\relax
  \def\bibnamefont#1{#1}\fi
\expandafter\ifx\csname bibfnamefont\endcsname\relax
  \def\bibfnamefont#1{#1}\fi
\expandafter\ifx\csname citenamefont\endcsname\relax
  \def\citenamefont#1{#1}\fi
\expandafter\ifx\csname url\endcsname\relax
  \def\url#1{\texttt{#1}}\fi
\expandafter\ifx\csname urlprefix\endcsname\relax\def\urlprefix{URL }\fi
\providecommand{\bibinfo}[2]{#2}
\providecommand{\eprint}[2][]{\url{#2}}

\bibitem[{\citenamefont{Aaij et~al.}(2016{\natexlab{a}})}]{LHCb-PAPER-2015-051}
\bibinfo{author}{\bibfnamefont{R.}~\bibnamefont{Aaij}} \bibnamefont{et~al.}
  (\bibinfo{collaboration}{LHCb collaboration}), \bibinfo{journal}{JHEP}
  \textbf{\bibinfo{volume}{02}}, \bibinfo{pages}{104}
  (\bibinfo{year}{2016}{\natexlab{a}}), \eprint{1512.04442}.

\bibitem[{\citenamefont{Dey}(2015)}]{Dey:2015rqa}
\bibinfo{author}{\bibfnamefont{B.}~\bibnamefont{Dey}}, \bibinfo{journal}{Phys.
  Rev.} \textbf{\bibinfo{volume}{D92}}, \bibinfo{pages}{033013}
  (\bibinfo{year}{2015}), \eprint{1505.02873}.

\bibitem[{\citenamefont{Aaij et~al.}(2016{\natexlab{b}})}]{lhcb_spd}
\bibinfo{author}{\bibfnamefont{R.}~\bibnamefont{Aaij}} \bibnamefont{et~al.}
  (\bibinfo{collaboration}{LHCb}) (\bibinfo{year}{2016}{\natexlab{b}}),
  \eprint{1609.04736}.

\bibitem[{\citenamefont{Chilikin et~al.}(2014)}]{Chilikin:2014bkk}
\bibinfo{author}{\bibfnamefont{K.}~\bibnamefont{Chilikin}} \bibnamefont{et~al.}
  (\bibinfo{collaboration}{Belle collaboration}), \bibinfo{journal}{Phys. Rev.}
  \textbf{\bibinfo{volume}{D90}}, \bibinfo{pages}{112009}
  (\bibinfo{year}{2014}), \eprint{1408.6457}.

\bibitem[{\citenamefont{Nishida et~al.}(2002)}]{Nishida:2002me}
\bibinfo{author}{\bibfnamefont{S.}~\bibnamefont{Nishida}} \bibnamefont{et~al.}
  (\bibinfo{collaboration}{Belle}), \bibinfo{journal}{Phys. Rev. Lett.}
  \textbf{\bibinfo{volume}{89}}, \bibinfo{pages}{231801}
  (\bibinfo{year}{2002}), \eprint{hep-ex/0205025}.

\bibitem[{\citenamefont{Aubert et~al.}(2004)}]{Aubert:2003zs}
\bibinfo{author}{\bibfnamefont{B.}~\bibnamefont{Aubert}} \bibnamefont{et~al.}
  (\bibinfo{collaboration}{BaBar}), \bibinfo{journal}{Phys. Rev.}
  \textbf{\bibinfo{volume}{D70}}, \bibinfo{pages}{091105}
  (\bibinfo{year}{2004}), \eprint{hep-ex/0409035}.

\bibitem[{\citenamefont{Lu and Wang}(2012)}]{Lu:2011jm}
\bibinfo{author}{\bibfnamefont{C.-D.} \bibnamefont{Lu}} \bibnamefont{and}
  \bibinfo{author}{\bibfnamefont{W.}~\bibnamefont{Wang}},
  \bibinfo{journal}{Phys. Rev.} \textbf{\bibinfo{volume}{D85}},
  \bibinfo{pages}{034014} (\bibinfo{year}{2012}), \eprint{1111.1513}.

\bibitem[{\citenamefont{Olive et~al.}(2014)}]{Agashe:2014kda}
\bibinfo{author}{\bibfnamefont{K.~A.} \bibnamefont{Olive}} \bibnamefont{et~al.}
  (\bibinfo{collaboration}{Particle Data Group}), \bibinfo{journal}{Chin.
  Phys.} \textbf{\bibinfo{volume}{C38}}, \bibinfo{pages}{090001}
  (\bibinfo{year}{2014}).

\bibitem[{\citenamefont{Sibidanov et~al.}(2013)}]{Sibidanov:2013rkk}
\bibinfo{author}{\bibfnamefont{A.}~\bibnamefont{Sibidanov}}
  \bibnamefont{et~al.} (\bibinfo{collaboration}{Belle}),
  \bibinfo{journal}{Phys. Rev.} \textbf{\bibinfo{volume}{D88}},
  \bibinfo{pages}{032005} (\bibinfo{year}{2013}), \eprint{1306.2781}.

\bibitem[{\citenamefont{Lees et~al.}(2016)}]{Lees:2015eya}
\bibinfo{author}{\bibfnamefont{J.~P.} \bibnamefont{Lees}} \bibnamefont{et~al.}
  (\bibinfo{collaboration}{BaBar}), \bibinfo{journal}{Phys. Rev. Lett.}
  \textbf{\bibinfo{volume}{116}}, \bibinfo{pages}{041801}
  (\bibinfo{year}{2016}), \eprint{1507.08303}.

\bibitem[{\citenamefont{Gilman and Singleton}(1990)}]{Gilman:1989uy}
\bibinfo{author}{\bibfnamefont{F.~J.} \bibnamefont{Gilman}} \bibnamefont{and}
  \bibinfo{author}{\bibfnamefont{R.~L.} \bibnamefont{Singleton}},
  \bibinfo{journal}{Phys. Rev.} \textbf{\bibinfo{volume}{D41}},
  \bibinfo{pages}{142} (\bibinfo{year}{1990}).

\bibitem[{\citenamefont{Korner and Schuler}(1990)}]{Korner:1989qb}
\bibinfo{author}{\bibfnamefont{J.~G.} \bibnamefont{Korner}} \bibnamefont{and}
  \bibinfo{author}{\bibfnamefont{G.~A.} \bibnamefont{Schuler}},
  \bibinfo{journal}{Z. Phys.} \textbf{\bibinfo{volume}{C46}},
  \bibinfo{pages}{93} (\bibinfo{year}{1990}).

\bibitem[{\citenamefont{Richman and Burchat}(1995)}]{Richman:1995wm}
\bibinfo{author}{\bibfnamefont{J.~D.} \bibnamefont{Richman}} \bibnamefont{and}
  \bibinfo{author}{\bibfnamefont{P.~R.} \bibnamefont{Burchat}},
  \bibinfo{journal}{Rev. Mod. Phys.} \textbf{\bibinfo{volume}{67}},
  \bibinfo{pages}{893} (\bibinfo{year}{1995}), \eprint{hep-ph/9508250}.

\bibitem[{\citenamefont{Aaij et~al.}(2013)}]{Aaij:2013iag}
\bibinfo{author}{\bibfnamefont{R.}~\bibnamefont{Aaij}} \bibnamefont{et~al.}
  (\bibinfo{collaboration}{LHCb}), \bibinfo{journal}{JHEP}
  \textbf{\bibinfo{volume}{08}}, \bibinfo{pages}{131} (\bibinfo{year}{2013}),
  \eprint{1304.6325}.

\bibitem[{\citenamefont{Aaij et~al.}(2016{\natexlab{c}})}]{Aaij:2015oid}
\bibinfo{author}{\bibfnamefont{R.}~\bibnamefont{Aaij}} \bibnamefont{et~al.}
  (\bibinfo{collaboration}{LHCb}), \bibinfo{journal}{JHEP}
  \textbf{\bibinfo{volume}{02}}, \bibinfo{pages}{104}
  (\bibinfo{year}{2016}{\natexlab{c}}), \eprint{1512.04442}.

\bibitem[{\citenamefont{Egede et~al.}(2010)\citenamefont{Egede, Hurth, Matias,
  Ramon, and Reece}}]{Egede:2010zc}
\bibinfo{author}{\bibfnamefont{U.}~\bibnamefont{Egede}},
  \bibinfo{author}{\bibfnamefont{T.}~\bibnamefont{Hurth}},
  \bibinfo{author}{\bibfnamefont{J.}~\bibnamefont{Matias}},
  \bibinfo{author}{\bibfnamefont{M.}~\bibnamefont{Ramon}}, \bibnamefont{and}
  \bibinfo{author}{\bibfnamefont{W.}~\bibnamefont{Reece}},
  \bibinfo{journal}{JHEP} \textbf{\bibinfo{volume}{10}}, \bibinfo{pages}{056}
  (\bibinfo{year}{2010}), \eprint{1005.0571}.

\bibitem[{\citenamefont{Becirevic et~al.}(2016)\citenamefont{Becirevic,
  Sumensari, and Zukanovich~Funchal}}]{Becirevic:2016zri}
\bibinfo{author}{\bibfnamefont{D.}~\bibnamefont{Becirevic}},
  \bibinfo{author}{\bibfnamefont{O.}~\bibnamefont{Sumensari}},
  \bibnamefont{and}
  \bibinfo{author}{\bibfnamefont{R.}~\bibnamefont{Zukanovich~Funchal}},
  \bibinfo{journal}{Eur. Phys. J.} \textbf{\bibinfo{volume}{C76}},
  \bibinfo{pages}{134} (\bibinfo{year}{2016}), \eprint{1602.00881}.

\bibitem[{\citenamefont{Gratrex et~al.}(2016)\citenamefont{Gratrex, Hopfer, and
  Zwicky}}]{Gratrex:2015hna}
\bibinfo{author}{\bibfnamefont{J.}~\bibnamefont{Gratrex}},
  \bibinfo{author}{\bibfnamefont{M.}~\bibnamefont{Hopfer}}, \bibnamefont{and}
  \bibinfo{author}{\bibfnamefont{R.}~\bibnamefont{Zwicky}},
  \bibinfo{journal}{Phys. Rev.} \textbf{\bibinfo{volume}{D93}},
  \bibinfo{pages}{054008} (\bibinfo{year}{2016}), \eprint{1506.03970}.

\bibitem[{\citenamefont{Hofer and Matias}(2015)}]{Hofer:2015kka}
\bibinfo{author}{\bibfnamefont{L.}~\bibnamefont{Hofer}} \bibnamefont{and}
  \bibinfo{author}{\bibfnamefont{J.}~\bibnamefont{Matias}},
  \bibinfo{journal}{JHEP} \textbf{\bibinfo{volume}{09}}, \bibinfo{pages}{104}
  (\bibinfo{year}{2015}), \eprint{1502.00920}.

\bibitem[{\citenamefont{Aubert et~al.}(2005)}]{Aubert:2004cp}
\bibinfo{author}{\bibfnamefont{B.}~\bibnamefont{Aubert}} \bibnamefont{et~al.}
  (\bibinfo{collaboration}{BaBar}), \bibinfo{journal}{Phys. Rev.}
  \textbf{\bibinfo{volume}{D71}}, \bibinfo{pages}{032005}
  (\bibinfo{year}{2005}), \eprint{hep-ex/0411016}.

\bibitem[{\citenamefont{Aaij et~al.}(2015)}]{Aaij:2014zsa}
\bibinfo{author}{\bibfnamefont{R.}~\bibnamefont{Aaij}} \bibnamefont{et~al.}
  (\bibinfo{collaboration}{LHCb}), \bibinfo{journal}{Phys. Rev. Lett.}
  \textbf{\bibinfo{volume}{114}}, \bibinfo{pages}{041801}
  (\bibinfo{year}{2015}), \eprint{1411.3104}.

\bibitem[{\citenamefont{Bernlochner et~al.}(2012)\citenamefont{Bernlochner,
  Ligeti, and Turczyk}}]{Bernlochner:2012bc}
\bibinfo{author}{\bibfnamefont{F.~U.} \bibnamefont{Bernlochner}},
  \bibinfo{author}{\bibfnamefont{Z.}~\bibnamefont{Ligeti}}, \bibnamefont{and}
  \bibinfo{author}{\bibfnamefont{S.}~\bibnamefont{Turczyk}},
  \bibinfo{journal}{Phys. Rev.} \textbf{\bibinfo{volume}{D85}},
  \bibinfo{pages}{094033} (\bibinfo{year}{2012}), \eprint{1202.1834}.

\end{thebibliography}

\end{document}